\documentclass[aps, prl, showpacs, reprint, superscriptaddress, twocolumn, tightenlines]{revtex4}
\usepackage{bm}
\usepackage{graphicx}
\usepackage{amssymb,amsmath,amsthm,color, bm}
\usepackage{microtype}
\usepackage{epsfig}

\usepackage{soul}

\newcommand{\beginsupplement}{%
        \setcounter{table}{0}
        \renewcommand{\thetable}{S\arabic{table}}%
        \setcounter{figure}{0}
        \renewcommand{\thefigure}{S\arabic{figure}}%
     }

\begin{document}
%\title{Mechanistic trail attachment leads to microcolony formation in \textit{Pseudomonas Aeruginosa}}
%\title{Self-organization of \textit{Pseudomonas Aeruginosa} due to mechanistic trail attachment}
\title{Multicellular self-organization of \textit{P. aeruginosa} due to interactions with secreted trails}

\author{Anatolij Gelimson}
\affiliation{Rudolf Peierls Centre for Theoretical Physics, University of Oxford, Oxford OX1 3NP, United Kingdom}

\author{Kun Zhao}
\affiliation{Key Laboratory of Systems Bioengineering, Ministry of Education, School of Chemical Engineering and Technology, Tianjin University, Tianjin, 300072, People's Republic of China}
\affiliation{Bioengineering Department, Chemistry \& Biochemistry Department, California Nano Systems Institute, UCLA, 90095-1600, Los Angeles, CA, USA}

\author{Calvin K. Lee}
\affiliation{Bioengineering Department, Chemistry \& Biochemistry Department, California Nano Systems Institute, UCLA, 90095-1600, Los Angeles, CA, USA}

\author{W. Till Kranz}
\affiliation{Rudolf Peierls Centre for Theoretical Physics, University of Oxford, Oxford OX1 3NP, United Kingdom}

\author{Gerard C. L. Wong}
\email{gclwong@seas.ucla.edu}
\affiliation{Bioengineering Department, Chemistry \& Biochemistry Department, California Nano Systems Institute, UCLA, 90095-1600, Los Angeles, CA, USA}

\author{Ramin Golestanian}
\email{ramin.golestanian@physics.ox.ac.uk}
\affiliation{Rudolf Peierls Centre for Theoretical Physics, University of Oxford, Oxford OX1 3NP, United Kingdom}

\date{\today}

\begin{abstract}
Guided movement in response to slowly diffusing polymeric trails provides a unique mechanism for self-organization of some microorganisms. To elucidate how this signaling route leads to microcolony formation, we experimentally probe the trajectory and orientation of \emph{Pseudomonas aeruginosa} that propel themselves on a surface using type IV pili motility appendages, which preferentially attach to deposited exopolysaccharides. We construct a stochastic model by analyzing single-bacterium trajectories, and show that the resulting theoretical prediction for the many-body behavior of the bacteria is in quantitative agreement with our experimental characterization of how cells explore the surface via a power law strategy. %This multi-scale description leads to ...
\end{abstract}


\maketitle

\begin{figure}[ht]
\centering
\includegraphics[width = 0.95 \columnwidth]{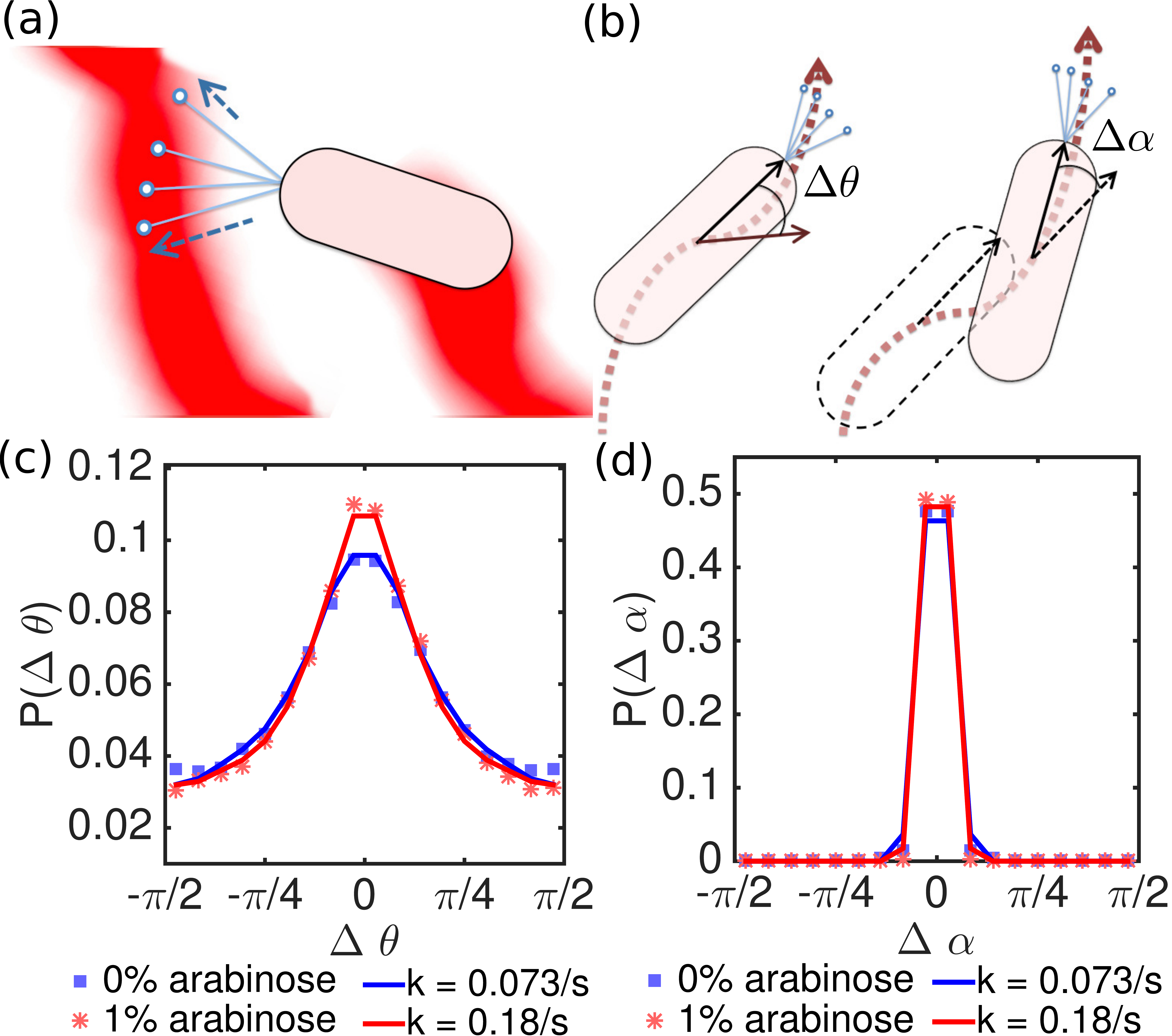}
\caption{(a) Bacterial model. Pili (light blue) preferentially attach to secreted Psl films on the surface (red) and pull the bacterium forward. (b) Schematic illustration for the angles $\Delta \theta$ (left) and $\Delta \alpha$ (right). $\Delta \theta$ is the angle between the current body orientation and the bacterial trajectory (dotted line), smoothed over $21$ points, which have been recorded every $3 \,\text{s}$, using a Savitsky-Golay filter of third order. $\Delta \alpha$ is the change in orientation between two timesteps of $3 \,\text{s}$. The resolution of an angle measurement is $\pm 5^{\circ}$ and about $400,000$ angle measurements were performed at an early stage in a sparse population. (c) shows the best fits between the experimental (dots) and the simulated distributions (solid lines) for $\Delta \theta$ (for different arabinose concentrations corresponding to different EPS secretion rates $k$), whereas (d) shows the corresponding $\Delta \alpha$ distributions. \label{fig:thetadist}}
\end{figure}

Chemotaxis -- biasing motility in response to chemicals -- is a fundamental organizing principle for guided movement in biological systems. Due to the long-range nature of the effect, it can strongly impact the dynamic organization of living \cite{Woodward:1995,Brenner:1998,benjacob+cohen00,liu2011sequential,zhang2012individual,taktikos2012collective,vergassola2007infotaxis,gelimson2015collective} and synthetic \cite{golestanian12,cohen+golestanian14,saha+golestanian14,holger} active matter. Chemotactic phenomena, with their implicit interpenetration between signal processing and migratory strategy \cite{sourjik+berg04,levine+rappel13}, have a diverse range, although they are often studied in the limit when the chemicals diffuse much faster than the cells. It is interesting to examine the opposite extreme, and investigate a highly-competitive, noisy system where every cell-secreted signal is long-lived and effectively non-diffusive. A model system for this scenario is {\em Pseudomonas aeruginosa}, which can secrete exopolysaccharides (EPS) \cite{ma2006analysis, ma2007pseudomonas, byrd2009genetic}. {\em Pseudomonas aeruginosa} moves on a 2D surface using the twitching motility mode via an EPS-binding motility appendage [Fig. \ref{fig:thetadist}(a)] known as a type IV pilus (TFP) \cite{burrows2012pseudomonas, maier2015bacteria, o1998flagellar, skerker2001direct, jin2011bacteria, Gibiansky:2013}. Recent work has shown that EPS play a key role in the clustering of cells into a microcolony \cite{Zhao:2013}, the first social step in bacterial biofilm formation, by facilitating pilus attachment. However, a quantitative understanding of the transition from single-bacteria motion to collective patterns is needed.
%little quantitative understanding exists, with no rigorous tests of the concept.

In the {\em P. aeruginosa} system, the EPS surface distribution that impacts a cell trajectory is dependent on the history of all other secreting cells that have traversed the surface, due to the long-lived nature of the Psl signal on the surface. This serves as a communal memory of particle trajectories, and leads to a coupling between cells that have large spatiotemporal separations. This provides the dominant mechanism for self-organization in the dilute regime, which is relevant for early stages of biofilm formation; in the dense regime, other effects such as excluded volume interaction \cite{Cates}, alignment \cite{Chate,Zhang,Perunai1,Peruani2,Shashi}, depletion interaction \cite{Herb2}, etc, will also impact the collective behavior of the system \cite{marchetti2013hydrodynamics}.

\begin{figure}
\centering
\includegraphics[width = \columnwidth]{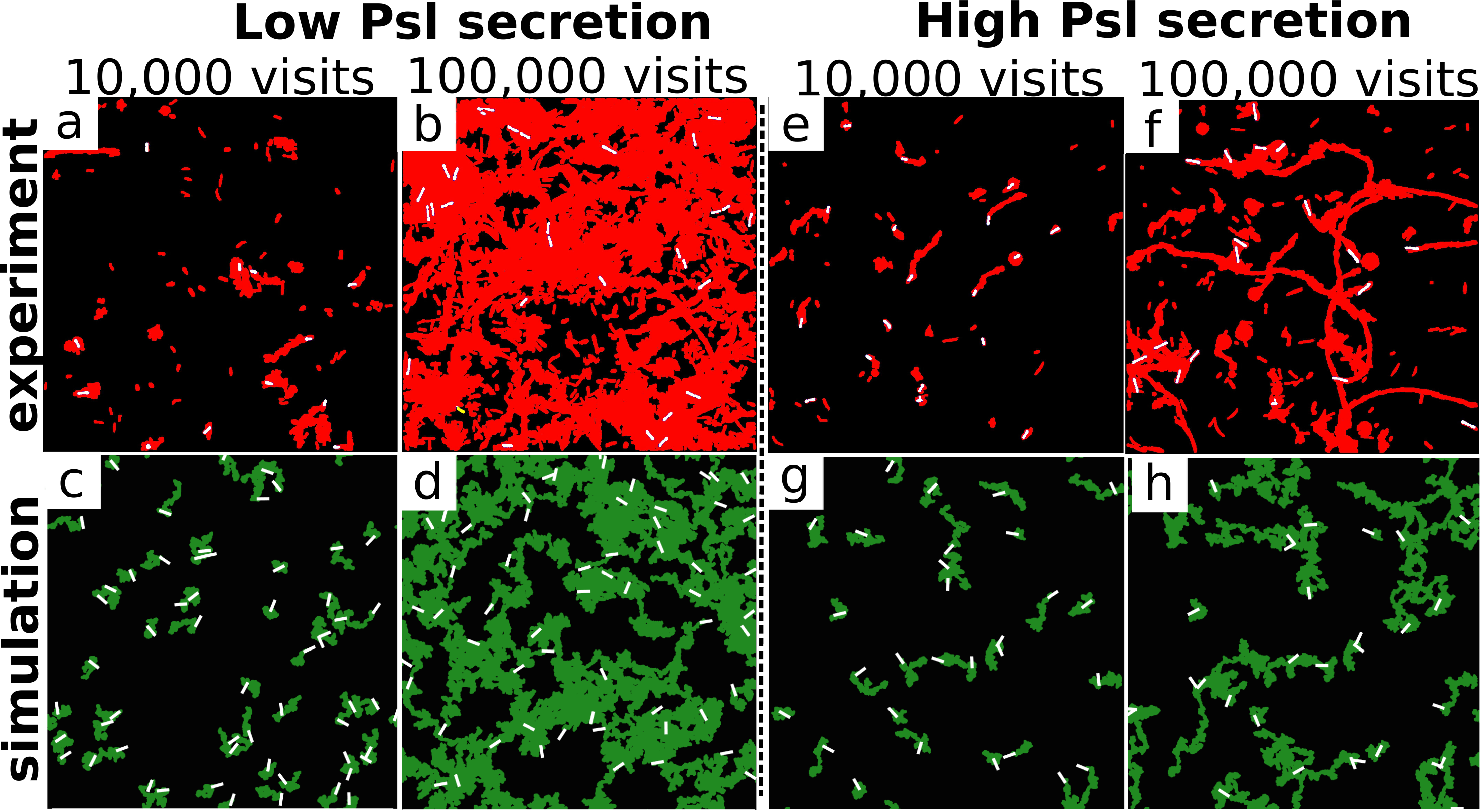}
\caption{Snapshots of bacterial positions (white) and surface areas visited over time from experiments (red) and simulations (green) for the case of low Psl deposition (a)--(d) and high Psl deposition (e)--(h). An enlarged version of this figure can be found in the supplement, Fig.~S1. (a)--(b) surface coverage for $\Delta$P$_{psl}$/P$_{BAD}$-{\it psl} under $0\%$ arabinose. Most of the surface has been visited after a total number of $S_{\text{visits}} =100,000$ bacterial visits. (c)--(d) Typical simulation outcome for the case of a low Psl deposition ($k = 0.073 \,\text{s}^{-1}$ corresponding to the best fit for $0 \%$ arabinose [see Fig. \ref{fig:thetadist}(c)--\ref{fig:thetadist}(d)]), starting from a random distribution of bacteria. In (d) bacteria explore most of the field. (e)--(f) show experimental data for $1\%$ arabinose. The visited area is much smaller and bacteria tend to aggregate more. (g)--(h) show a typical simulation outcome of high Psl deposition ($k = 0.18 \,\text{s}^{-1}$, corresponding to $1 \%$ arabinose), starting from a random distribution. As shown in (h) bacteria aggregate in a small, Psl-rich surface area, making it more attractive for others by secreting more Psl there.} \label{fig:snapshots}
\end{figure}

\begin{figure}
\centering
\includegraphics[width = 0.9 \columnwidth]{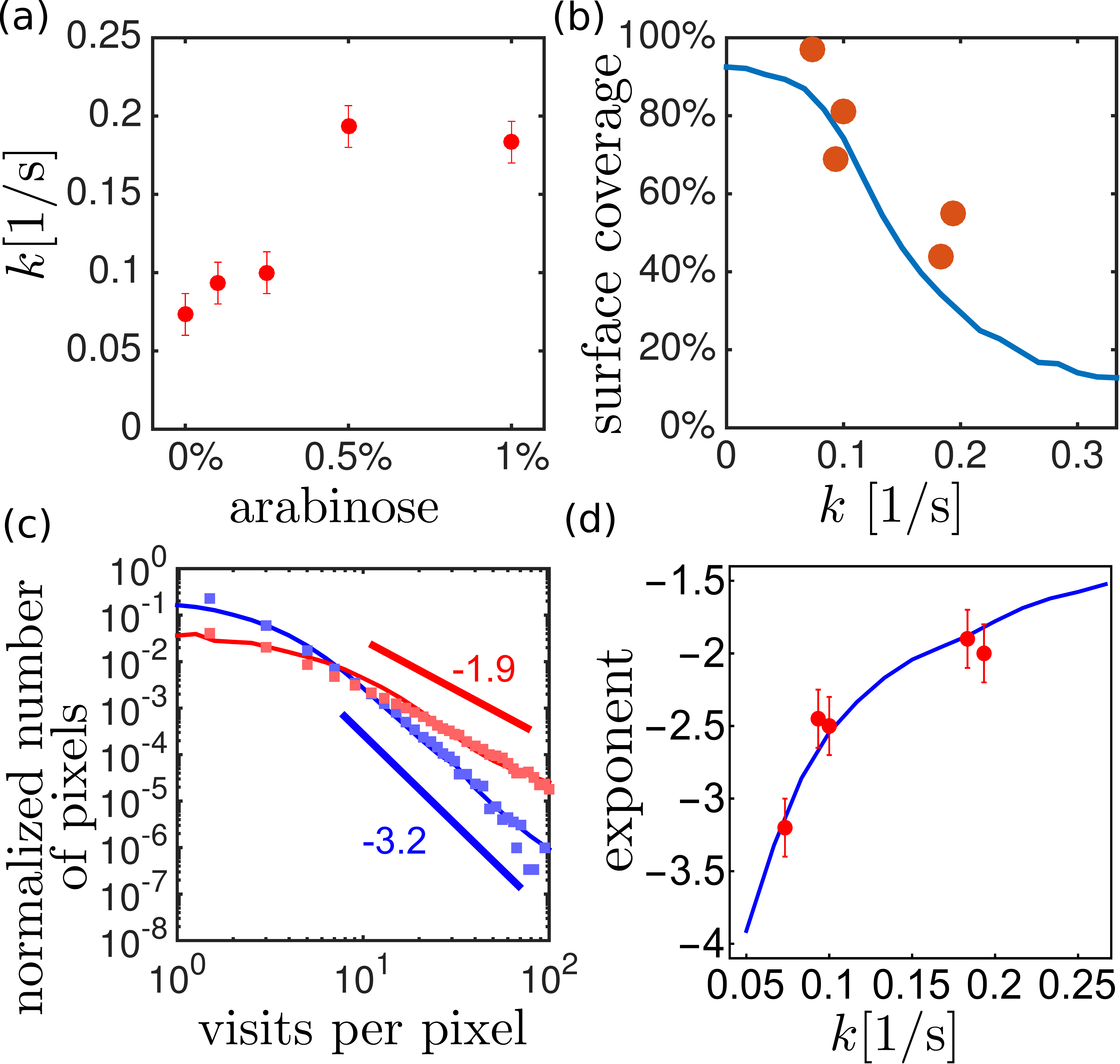}
\caption{(a) Experimental arabinose concentrations $C_{\text{ara}}=(0\%, 0.1\%, 0.25 \%, 0.5 \%, 1 \%)$ and the corresponding theoretical values for the Psl secretion $k = (0.073, 0.093, 0.1, 0.19, 0.18) \,\text{s}^{-1}$, obtained from the distribution for $\Delta \theta$ and $\Delta \alpha$ at early stages [see Figs. \ref{fig:thetadist}(b)--\ref{fig:thetadist}(d)]. (b)--(d) Comparison between experiments (dots) and theory (lines) for the collective behavior of $\Delta$P$_{psl}$/P$_{BAD}$-{\it psl}. No additional fitting was required to match the experimental results. (b) Surface percentage visited by a colony of $30$ bacteria after $13.8$ h ($S_{\text{visits}}  = 500,000$ total visits) as a function of Psl deposition rate $k$. Red dots show experimental data for $0 \%$, $0.1 \%$, $0.25 \%$, $0.5 \%$ and $1 \%$ arabinose [for the corresponding $k$ values see (a)], with an error of $\pm 10 \%$ for $0 \%$, $0.1 \%$ and $1 \%$ arabinose. For $0.25 \%$ and $0.5 \%$ the error could not be determined due to a low sample size. (c) The result of the rich-get-richer mechanism is a power-law distribution in the frequency of visits per pixel, shown for low (blue) and high Psl secretion (red). Blue: low Psl secretion, $0 \%$ arabinose or $k= 0.073\,\text{s}^{-1}$. Red: high Psl secretion rate at $1 \%$ arabinose or $k= 0.18\,\text{s}^{-1}$. The power-law distributions become more hierarchical (slower decay) with increasing Psl. (d) Comparison between experimental (red dots with error bars) and simulated (blue line) power-law exponents. The estimated error of the exponents is $\pm 0.1$ for simulations and $\pm 0.2$ for experiments. \label{fig:collective}}
\end{figure}

Here, we present a quantitative description of how the microscopic motion of a single EPS-depositing bacterium directly leads to clustering and microcolony formation in the collective case. We (1) experimentally probe single-bacterium dynamics by measuring the trajectories and bacterial orientations [see Fig. \ref{fig:thetadist}], (2) theoretically derive a microscopic model for pili-driven motility, (3) extract all parameters on a single-particle level from experiments, and (4) show that the same model quantitatively explains experimentally measured observables in the early stages of biofilm formation, without any additional adjustments [see Figs. \ref{fig:snapshots} and \ref{fig:collective}]. Our multi-scale description provides a quantitative understanding of how cells interact with their system of mutually secreted trails, and spontaneously explore the surface via a power law migratory strategy. A crucial consequence of TFP sensing is the emergence of novel features in the orientational dynamics \cite{till}, which we fully characterize both experimentally and theoretically.

%quantitative description of this phenomenon, which has the following three components: (1) we experimentally probe single-bacterium dynamics by measuring the trajectories and bacterial orientations [see Fig. \ref{fig:thetadist}], (2) we next build a microscopic model for pili-driven motility describing the dynamics of position and orientation, and extract its parameters from the single-bacterium experiments, and (3) we then perform a many-body simulation of the resulting model, and show that our results are in quantitative agreement with the relevant experimentally measured collective properties without a need for adjustments or additional fitting [see Fig. \ref{fig:snapshots}].
%not just a gradient dependent velocity but also a seemingly counter-intuitive gradient dependent perpendicular alignment term \cite{till}, which stabilizes the pattern formation.

%\paragraph{Single-Bacterium Behavior---}
In our experiments we used the {\em P. aeruginosa} PAO1 \cite{holloway1955genetic} strain $\Delta$P$_{psl}$/P$_{BAD}$-{\it psl} \cite{ma2006analysis}, which is a strain with inducible Psl secretion controlled by arabinose concentrations. For the detailed experimental information, see methods in Ref. \cite{Zhao:2013}. An overnight bacteria culture in FAB medium \cite{heydorn2000experimental} supplemented with 30 mM glutamate, was diluted and injected into a flow cell \cite{Zhao:2013, crusz2012flowcell}. The FAB medium with 0.6 mM glutamate was continuously pumped through the flow chamber using a syringe pump with a flow rate of 3 ml/hour at 30$^{\circ}$C. Different amounts of arabinose were added into the medium to control the production of Psl. Bright-field images were taken every 3 seconds by an EMCCD camera on an Olympus IX81 microscope equipped with a Zero Drift autofocus system. The image size is $67 \times 67 \mu \text{m}^2$ ($1024 \times 1024 \text{px}^2$, where px $\equiv$ pixel size $= 0.06536 \,\mu \text{m}$). A typical data set has about 14,000$-$20,000 frames with a total of up to 1,000,000 snapshots of single bacteria. By analyzing those data sets using a parallel cell-tracking algorithm \cite{Zhao:2013}, which extracts the full motility history of each tracked bacterium in the field of view,
%we investigated the effects of pili attachment Psl deposition on the motion of single bacterium, and
and by analyzing the time dependence of fluorescently labeled Psl trails, we investigate how these trails impact bacterial motility. For better comparability of samples with different numbers of bacteria on the surface, we have measured time in terms of total bacterial visits, i.e. the sum of the number of bacteria in all frames $S_{\text{visits}}  = \sum s_i$, where $s_i$ is the number of bacteria present in frame $i$ \cite{Zhao:2013}.

The distributions of the two angles $\Delta \theta$ and $\Delta \alpha$ defined in Fig. \ref{fig:thetadist}(b) are shown in Figs. \ref{fig:thetadist}(c)--\ref{fig:thetadist}(d) for $0\%$ and $1\%$ arabinose. Figure \ref{fig:thetadist}(d) shows the distributions of the change in bacterial body orientation $\Delta \alpha$ within a $3\,\text{s}$ time frame; they appear to be very narrow and mostly independent of the strains in the experiment, which is due to a relatively small rotational diffusion (Fig. S2(b) \cite{Note1}). On the other hand, the distribution of the angle between the bacterial trajectory and the body orientation, $\Delta \theta$, is wide. Importantly, with increasing Psl deposition the $\Delta \theta$ distribution narrows down [see Fig. \ref{fig:thetadist}(c)], showing that Psl stabilizes the bacterial trajectory and increases the correlation between orientation of the bacterial body and the moving direction.

%\paragraph{Collective Behavior---}
We consider a mechanistic model for a rod-like bacterium that pulls itself forward on a surface using pili, taking into account the influence of a bacteria-secreted polysaccharide film on the attachment/detachment of pili, as well as on pulling forces; see Supplemental Material \cite{Note1}.  The model is an extension of the model in \cite{till}, but here we will specifically focus on \emph{P. aeruginosa} to better understand the experimental results for single-bacterium motion and collective behavior. Each bacterium (labeled) $a$ is assigned a position $\bm{r}_a$ and an orientation $\hat{\bm{n}}_a = (\cos \varphi_a(t), \sin \varphi_a(t))$ [we can also define a perpendicular unit vector $\hat{\bm{n}}_{\perp a} = (-\sin \varphi_a(t), \cos \varphi_a(t))$]. The bacteria secrete a Psl trail, to which pili can preferentially attach. The attached pili then pull the bacteria forward. The Psl trail has a finite width $\delta$, which we assume to be of the same order as the width of a bacterium. As a result of Psl-dependent attachment/detachment and pulling, pili can propel the bacterium along $\hat{\bm{n}}_a$, but also along the Psl gradient $\bm{\nabla} \psi$, depending on the distribution of pili.  In addition, the Psl-dependent pulling will result in an alignment term towards the gradient $\bm{\nabla} \psi$ \cite{till}. Finally, due to fluctuations in attachment/detachment of bacterial pili, the overall pili pulling force acting on a bacterium will fluctuate; this will introduce noise in the position and orientation Langevin equations, which we approximate as Gaussian. Overall, the equations of motion for the many-body dynamics are constructed as follows \cite{Note1}
\begin{math}
\frac{d \bm{r}_a}{d t} \;= \;A(\psi) \; (\bm{\nabla} \psi \cdot \hat{\bm{n}}_{\perp a}) \,\hat{\bm{n}}_{\perp a} + B(\psi) \; (\bm{\nabla} \psi \cdot \hat{\bm{n}}_a) \,\hat{\bm{n}}_a + \, v(\psi) \, \hat{\bm{n}}_a + \sqrt{ 2 D_\parallel(\psi)} \; \eta^\parallel_a \, \hat{\bm{n}}_a + \sqrt{2 D_\perp(\psi)} \; \eta^{\perp}_a \, \hat{\bm{n}}_{\perp a},
\end{math}
\begin{math}
\frac{d \hat{\bm{n}}_a}{d t} =  -  \chi(\psi) \, \hat{\bm{n}}_a \times[\hat{\bm{n}}_a \times \bm{\nabla} \psi] +  \sqrt{2 D_r(\psi)} \; \eta^{\perp}_a \, \hat{\bm{n}}_{\perp a},
\end{math}
and
\begin{math}
\partial_t \psi(\bm{r}, t) = k \sum_a \frac{1}{2\pi\delta^2}\; e^{-(\bm{r}-\bm{r}_a)^2/2\delta^2}.
\end{math}
The coefficients are related, and can be written in terms of two independent ones (we choose $D_r$ and $v$):
%
%\begin{subequations}
\begin{math}
\chi = \left[\frac{\gamma_{\parallel} \ell \ell_p (1-c_2)}{2 \gamma_{\text{rot}} c_1} \right] \partial_\psi v,
\end{math}
\begin{math}
D_\parallel = \left[\frac{4 \gamma_{\text{rot}}^2 c_2}{ (1-c_2) \gamma_\parallel^2 \ell^2}\right] D_r,
\end{math}
\begin{math}
D_\perp = \left(\frac{2 \gamma_{\text{rot}} }{\gamma_\perp \ell} \right)^2 D_r,
\end{math}
\begin{math}
A = \left[\frac{\ell_p (1-c_2) \gamma_\parallel}{c_1 \gamma_\perp} \right] \partial_\psi v,
\end{math}
and
\begin{math}
B = \left[\frac{c_2 \ell_p}{c_1}\right] \partial_\psi v.
\end{math}
%\end{subequations}
%
Here $\ell$ is the length of a bacterium, $\ell_p$ is the length of a pilus, and $\gamma_\parallel$, $\gamma_\perp$, $\gamma_{\text{rot}}$ are the friction coefficients for the translation parallel/perpendicular to $\hat{\bm{n}}_a$ and for rotational motion. $c_1 = \langle \cos \vartheta \rangle$ and $c_2=\langle \cos^2 \vartheta \rangle$ denote averages involving the distribution of pili angles $\vartheta$ relative to the bacterial body axis [see Fig. \ref{fig:thetadist}(a)]. The noise terms, $\eta^{\perp}_a$ and $\eta^{\parallel}_a$, are independent Gaussian white noise of unit strength \cite{Note1}. We note that the particular geometry for the pili has made the {\em orientational noise} identical to the {\em transverse translational noise}; see Eq. (2). This is a significant feature of our model. Our choice of a Gaussian shape for the bacterial Psl trail is motivated by the assumption that Psl just secreted from the bacteria will initially diffuse but then the diffusion will come to an effective stop due to stickiness of Psl. Similar models have been applied in previous studies \cite{Couzin:2003}.

Experimentally, the length of a bacterium has been determined as $\ell \approx 2 \,\mu \text{m}$, and the length of a pilus $\ell_p$ is of the order of the body length. The apparent width of a bacterium has been determined as approximately $0.5 \, \mu \text{m}$. Since the bacteria are approximately rod-like, we estimate $\gamma_{\perp}/\gamma_{\text{rot}} \approx 5/\ell^2$ \cite{tirado1984}. In addition, the anisotropy in friction (perpendicular and parallel to the long axis) is $(\gamma_\perp-\gamma_\parallel)/\gamma_\parallel \approx 0.2$ \cite{tirado1984} % and hence \anote{ $D_\parallel / D_\perp \approx 1.5 \, c_2/(1-c_2)$,
where all values have been rounded. Experimentally, in the absence of EPS secretion we find a diffusion constant $D_{\text{total} 0} = 1.1 \times10^{-2} \mu \text{m}^2\,\text{s}^{-1}$ (Fig. S2(a) \cite{Note1}), which is the result of parallel and perpendicular diffusion, $D_{\text{total}0} = D_{\parallel 0} + D_{\perp 0} = D_{\perp 0} [1 + c_2 \gamma_\perp^2 / (1-c_2) \gamma_\parallel^2]$. With $D_{r} \approx 3.6 \times10^{-3} \,\text{s}^{-1}$ for $0 \%$ arabinose (Fig. S2(b) \cite{Note1}) we get $D_{\perp} = 2.3\times10^{-3} \mu \text{m}^2\,\text{s}^{-1}$ and $c_2 = \langle \cos^2 \vartheta \rangle \approx 0.7$. As an approximation, in our simulations we will set $c_1 = \langle \cos \vartheta \rangle \approx \sqrt{c_2} = 0.8$. The characteristic width of the Gaussian Psl trail has been chosen as $\delta = 0.25 \, \mu \text{m}$, half the width of the bacterium.

Our microscopic model shows that the dependencies of $v$ and $D_r$ on $\psi$ are controlled by the pili pulling force $f$, the friction coefficients $\gamma_\parallel$ and $\gamma_{\rm rot}$, the attachment rate $\lambda$, and the detachment rate $\mu$: $v(\psi) \propto f \lambda/[\gamma_\parallel (\lambda+\mu)]$ and $D_r(\psi) \propto f^2 \lambda \mu/[\gamma_{\rm rot}^2 (\lambda+\mu)^3]$ (see Supplemental Material \cite{Note1}). While in principle all of these quantities depend on $\psi$, we can make some intuitive assumptions about the general trends; e.g. the friction coefficients increase with $\psi$, and better attachment to Psl implies the attachment rate increases, and the detachment rate decreases, as $\psi$ is increased. This, in turn, suggests that by extracting information about the dependence of $v$ and $D_r$ on $\psi$ from experiments, we can gather information about the above mentioned components. If the Psl concentrations are small, we can Taylor-approximate $v\approx v_0 + v_1 \psi$ and $D_r \approx D_{r 0} + D_{r 1} \psi$.
%The signs in the Taylor expansions account for the intuition that the bacterial velocity will to a first approximation increase with $\psi$ due to better attachment and pili pulling. Furthermore, better attachment to the surface will decrease random behaviour and hence we expect $D_\perp(\psi)$ to decrease with increasing $\psi$. This intuition is corroborated by our microscopic model, which relates these quantities to the attachment rate $\lambda(\psi)$, and the detachment rate $\mu(\psi)$, as follows: $v(\psi) \propto \lambda(\psi)/[\lambda(\psi)+\mu(\psi)]$ and $D_\perp(\psi) \propto \lambda(\psi)\mu(\psi)/[\lambda(\psi)+\mu(\psi)]^2$ (see Supplemental Material \cite{Note1}). The dependence can lead to the anticipated behavior if by increasing $\psi$ the attachment rate increases and the detachment rate decreases.
For a mutant without any EPS secretion, the experimental data for the translational MSD shows no propulsive component of the motion at the shortest timescale that we have resolved (Fig. S2(a) \cite{Note1}), and hence we conclude that in the absence of Psl the velocity component $v_0 \approx 0$. We therefore model the velocity as $v = v_1 \psi$. This also implies that $A(\psi)$, $B(\psi)$ and $\chi(\psi)$ are larger than zero for non-zero EPS secretion. In addition, noting that the rotational diffusion $D_r$ varies only weakly between $0 \%$ and $1 \%$ arabinose, we set $D_{r 1} \approx 0$, which also introduces non-zero translational diffusion constants $D_\parallel$ and $D_\perp$.

We performed Brownian dynamics simulations of the dynamical equations with the above choices for the coefficients and parameters. Note that the Psl deposition strength $k$ always appears in combination with $v_1$ and hence does not constitute an additional parameter. This leaves us with the one parameter $k v_1$ to be determined. In our simulations we only varied the Psl deposition $k$ and set $v_1 = 1 \text{px}^3/\text{frame} = 9.3 \times 10^{-5} \mu \text{m}^3\,\text{s}^{-1}$. The parameters in the numerical simulations were chosen such that they correspond to the experimental parameters. The unit of length was $1$ px $= 0.06536 \,\mu \text{m}$ and the unit of time was 1 frame $= 3\,\text{s}$. For all our simulations, we regarded the above parameters as characteristic constants of the bacteria and only varied $k$ between simulation runs to emulate a situation where the amount of Psl produced depends on an external factor (like the arabinose concentration on the substrate for the $\Delta$P$_{psl}$/P$_{BAD}$-{\it psl} mutant).

Figures \ref{fig:thetadist}(c)--\ref{fig:thetadist}(d) show the angle distributions for $\Delta \theta$ and $\Delta \alpha$ from experiments and simulations. The qualitative characteristics from the experiments, in particular the very narrow distribution for $\Delta \alpha$, the wide distribution for $\Delta \theta$ and the narrowing down of $\Delta \theta$ with increasing Psl deposition have been reproduced. As highlighted in the Supplemental Material \cite{Note1}, the translational and alignment terms proportional to $\bm{\nabla} \psi$ in Eqs. (1) and (2) confine the bacteria to the trail center and enable them to follow trails of other bacteria, thus increasing the correlation between the trail and body orientation at long times. Comparison between simulations and experiments suggests that the experimental data for low Psl deposition ($0\%$ arabinose) best correspond to $k= 0.073,\text{s}^{-1}$ [Fig. \ref{fig:thetadist}(c)], whereas the data for high Psl deposition ($1\%$ arabinose concentration) are best matched by $k= 0.18\,\text{s}^{-1}$ [Fig. \ref{fig:thetadist}(d)]. The corresponding $k$ values for other arabinose concentrations are shown in Fig. \ref{fig:collective}(a).

On a larger scale, studies using single-particle tracking have shown that the Type IV-pili-mediated twitching motility of {\em P. aeruginosa} plays an essential role in the early stages of biofilm formation due to complex interactions with secreted Psl. The result of this is a ``rich get richer mechanism'' which helps {\em P. aeruginosa} accumulate at just a few sites, whereas other sites are not visited at all or visited infrequently. To put our microscopic model to stringent test, we compare the long-time prediction of the simulation with the corresponding results from our experiments without any additional fitting or adjustments. We find that the simulations predict exactly the same behavior as we observe experimentally, including the power law distribution of visited sites and the percentage of surface coverage; see Figs. \ref{fig:collective}(b)--(d). Figures \ref{fig:snapshots}(a)--\ref{fig:snapshots}(h) show a typical Psl field $\psi(\bm{r}, t)$ over a long period of time. In Figs.~\ref{fig:snapshots}(a)--\ref{fig:snapshots}(d) the Psl deposition strength is very small, which makes the bacterial behavior essentially diffusive for long times. Therefore in the long run the spatial distribution of Psl is relatively uniform since the bacteria are not attracted to any particular sites. This is reflected in a high surface coverage over time [Fig.~\ref{fig:collective}(b)], as well as a relatively small power-law exponent for the distribution, Figs.~\ref{fig:collective}(c)--\ref{fig:collective}(d).

On the other hand, in Figs.~\ref{fig:snapshots}(e)--\ref{fig:snapshots}(h), which show the Psl field $\psi$ for a large Psl deposition rate, sites are preferentially visited where bacteria have already been previously, due to the strong interaction with the Psl field. As a result, bacteria visit a smaller fraction of sites; Fig.~\ref{fig:collective}(b). Analyzing the distribution of total bacterial visits to a given pixel one observes a power-law behavior when plotting the total visits per pixel against the number of pixels with this visit frequency. For an increasing Psl deposition rate, the exponent increases, which implies a less egalitarian visit distribution, i.e. many sites with very few bacterial visits and a few sites where bacteria aggregate and spend a long period of time. This leads to a hierarchical Psl distribution, with a larger power-law exponent; Figs.~\ref{fig:collective}(c)--\ref{fig:collective}(d). The distribution of visits is directly related to the distribution of Psl concentrations per pixel \cite{Zhao:2013}.

The distribution of bacterial visits to surface sites (and consequently the Psl concentration) has the hallmarks of a Pareto-type power-law distribution, similar to wealth distributions in capitalist economies \cite{newman2005power, Zhao:2013}. Moreover, experiments with the $\Delta$P$_{psl}$/P$_{BAD}$-{\it psl} mutant, which produces Psl depending on the arabinose concentration on the substrate (varied between $0\%$ and $1\%$), suggest that the corresponding power-law exponents change from $-3.1$ to $-1.8$, making the Psl and surface visit distributions more hierarchical with increasing Psl deposition \cite{Zhao:2013}. Theoretical and experimental exponents are in good quantitative agreement, as seen in Fig. \ref{fig:collective}(d). The simulations for Figs. \ref{fig:collective}(b)--\ref{fig:collective}(d) were performed on a field with $30$ bacteria for $16667$ frames, i.e. a total of $S_{\text{visits}} =500,000$.%, which is the equivalent to the typical measurement time of $13.8$ h in experiments.

%\paragraph{Conclusions ---}

In summary, based on microscopic interactions observed in {\em P. aeruginosa} experiments, we have shown that the microcolony formation can be a direct consequence of trail deposition and pili-mediated motility, without any active signal processing by the bacteria. This leads to efficient surface exploration and self-organization.
%We found orientational alignment with the gradient to be crucial for explaining social trail-following and the experimental measurements for the angle distributions. %Orientational alignment facilitates the following of existing trails and stabilizes bacterial trajectories which would otherwise be disrupted by orientational noise (such as %those from twitching motility \cite{jin2011bacteria}) }.
We have found that the combination of translational and orientational gradient response facilitates social trail-following and thus quantitatively explains crucial effects at the early stages of microcolony formation. We expect these trail-induced motility effects to strongly modify the complex motility from competitive deployment of multiple type IV pili \cite{Berenike}. Our results have a broad range of impact, ranging from self-guided active matter to bacterial biofilm microbiology to embryonic development.

\begin{acknowledgments}
This work was supported by the Human Frontier Science Program RGP0061/2013, the Ernst Ludwig Ehrlich Studienwerk (A.G.), the Tianjin Municipal Natural Science Foundation grant 15JCZDJC41100 (K.Z.), and the Office of Naval Research grant N000141410051 (G.C.L.W.). We acknowledge the COST Action MP1305 ``Flowing Matter''.
%We are grateful to ... for discussions.
\end{acknowledgments}

\onecolumngrid
\appendix

\beginsupplement

\section{\fontsize{12}{15} \sc \selectfont Supplemental Material} 

\renewcommand{\theequation}{S\arabic{equation}}

\section{\fontsize{12}{15} \sc \selectfont I. FIG. 2: Bacterial Positions and Visited Surface Areas (Enlarged)}

\begin{figure}[h]
\centering
\includegraphics[width = \textwidth]{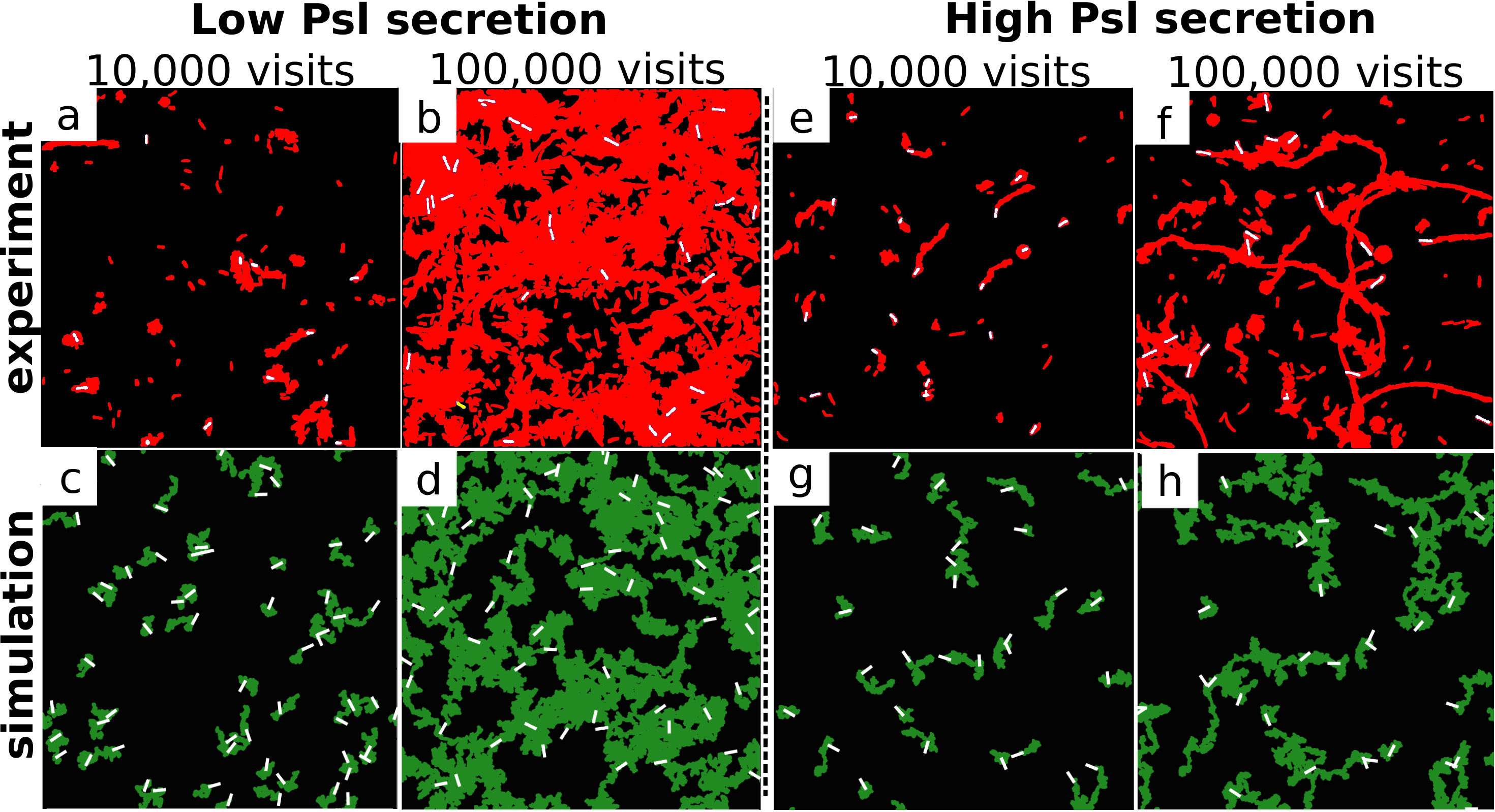}
\caption{Enlarged version of Fig. 2. Snapshots of bacterial positions (white) and surface areas visited over time from experiments (red) and simulations (green) for the case of low Psl deposition (a)--(d) and high Psl deposition (e)--(h). (a)--(b) surface coverage for $\Delta$P$_{psl}$/P$_{BAD}$-{\it psl} under $0\%$ arabinose. Most of the surface has been visited after a total number of $S_{\text{visits}} =100,000$ bacterial visits. (c)--(d) Typical simulation outcome for the case of a low Psl deposition ($k = 0.073 \,\text{s}^{-1}$ corresponding to the best fit for $0 \%$ arabinose [see Fig. 1(c)--1(d)]), starting from a random distribution of bacteria. In (d) bacteria explore most of the field. (e)--(f) show experimental data for $1\%$ arabinose. The visited area is much smaller and bacteria tend to aggregate more. (g)--(h) show a typical simulation outcome of high Psl deposition ($k = 0.18 \,\text{s}^{-1}$, corresponding to $1 \%$ arabinose), starting from a random distribution. As shown in (h) bacteria aggregate in a small, Psl-rich surface area, making it more attractive for others by secreting more Psl there.  } \label{fig:snapshots-large}
\end{figure}

\section{\fontsize{12}{15} \sc \selectfont II. Microscopic Derivation of the Stochastic Equations of Motion}

Here we describe the detailed derivation of the stochastic equations of motion for the bacteria. The theoretical formulation is based on a microscopic motility model that takes into account stochastic attachment and detachment of pili, as well as friction coefficients and pili contraction forces that depend on the amount of Psl.

\begin{figure}[b]
%\centering
\includegraphics[width =  0.3 \columnwidth]{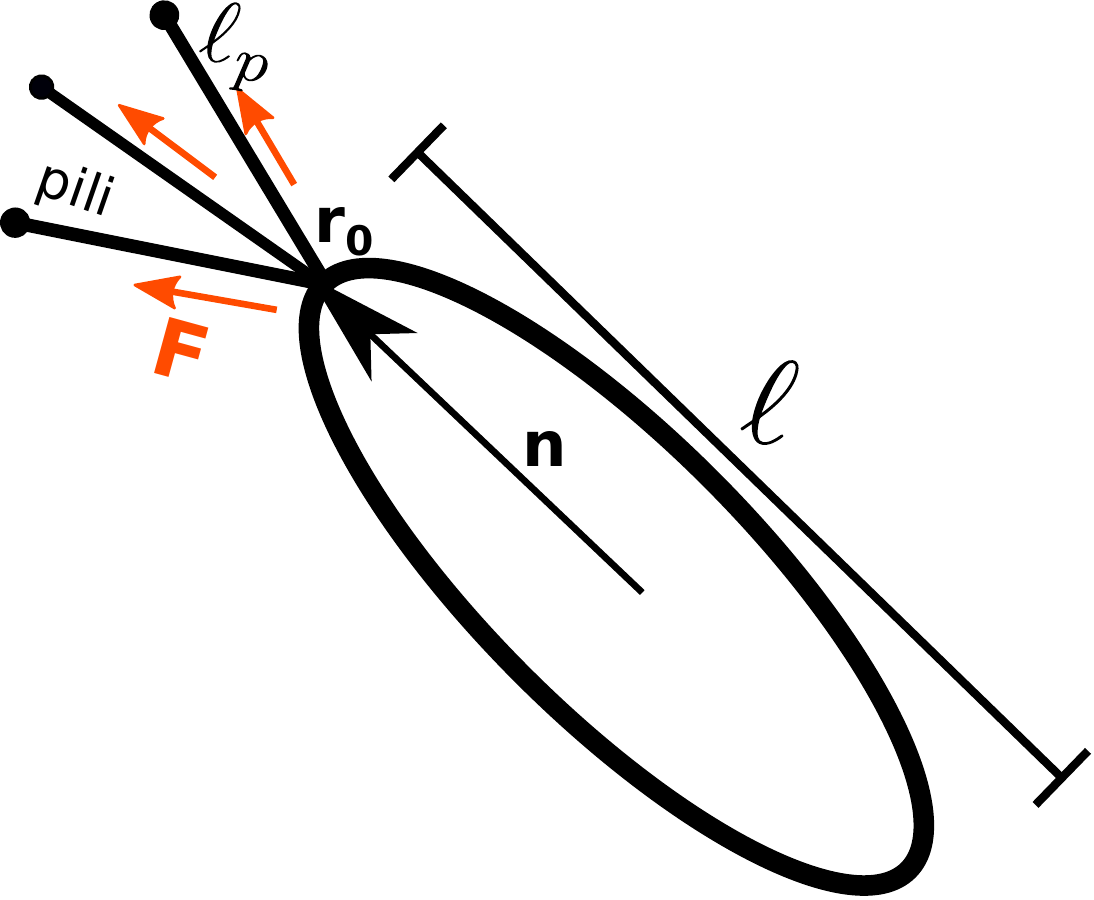}
\caption{Microscopic model: the bacterium has type-IV pili of length $\ell_p$, which can attach to a surface and pull the bacterium forward with a generically Psl-dependent force. Originating at a bacterial pole $\bm{r}_0$, the pili point in different directions $\hat{\bm{e}}_i, \; i = 1,\dots,N$, such that the pili tips are located at the positions $\bm{r}_0 + \ell_p \hat{\bm{e}}_i$. A pilus pulling force $f(\psi_i) \hat{\bm{e}}_i$ on the bacterial tip $\bm{r}_0$ is only generated if the pilus is attached to the surface and in general the attachment and detachment rates $\lambda(\psi_i)$ and $\mu(\psi_i)$ will be dependent on the Psl concentration $\psi_i = \psi( \bm{r}_0 + \ell_p \hat{\bm{e}}_i )$ at the end of pilus $i$ \label{model-schematics}}
\end{figure}

We regard a bacterium of length $\ell$ (Fig. S\ref{model-schematics}). We will assume that each of the $N$ type-IV pili located at one tip (or at both tips) of a bacterium will attach to/detach from the surface and pull in a Psl-dependent way. The direction unit vectors of the pili $\bm{e}_i, \; i = 1,\dots, N$ are assumed to originate from the bacterial tip $\bm{r}_0$. One could consider variants of this model where the pili distribution is different. A single pilus tip, located at $\bm{r}_0 + \ell_p \hat{\bm{e}}_i$, will randomly attach to the surface with a Psl-dependent rate $\lambda(\psi_i)$, detach with a rate $\mu(\psi_i)$ and pull in its direction with a force $f(\psi_i)$ if it is attached to the surface. Here, $\psi_i$ is the Psl concentration at the tip of pilus $i$, $\psi_i = \psi( \bm{r}_0 + \ell_p \hat{\bm{e}}_i )$. For each pilus $i$ we define $\Theta_i = 1$ if the pilus is attached and $\Theta_i = 0$ if it is detached. $\Theta_i$ will have a mean
\begin{equation}
\bar{\Theta} = \langle \Theta_i \rangle = \frac{\lambda(\psi)}{\lambda(\psi) + \mu(\psi)},
\end{equation}
and a variance
\begin{equation}
\sigma^2 = \langle (\Theta_i - \bar{\Theta})^2 \rangle = \frac{\lambda(\psi)\mu(\psi)}{[\lambda(\psi) + \mu(\psi)]^2}.
\end{equation}
The total force exerted at the bacterial tip will then be
\begin{equation}
\bm{F} =  \sum_i \hat{\bm{e}}_i f(\psi_i) \Theta_i.
\end{equation}

Rewriting $\Theta_i = \bar{\Theta} + \delta \Theta_i$, and Taylor-expanding
$$\psi_i = \psi( \bm{r}_0 + \ell_p \hat{\bm{e}}_i ) \approx \psi( \bm{r}_0) + \ell_p ( \bm{\nabla} \left. \psi \right|_{\bm{r}_0} \cdot \hat{\bm{e}}_i)$$
we get
\begin{equation}
%\begin{split}
\bm{F} =  \sum_i \hat{\bm{e}}_i f(\psi) \bar{\Theta}(\psi) + \sum_i \hat{\bm{e}}_i \ell_p (\bm{\nabla} \psi \cdot \hat{\bm{e}}_i) \partial_\psi( f \bar{\Theta})+ \sum_i \hat{\bm{e}}_i f(\psi_i) \delta \Theta_i.
%\end{split}
\end{equation}

We can express the pili orientation vectors in terms of the bacterial body orientation $\hat{\bm{n}}$ and the orientation $\hat{\bm{n}}_\perp = \hat{\bm{e}}_z \times \hat{\bm{n}}$  orthogonal to it as $\hat{\bm{e}}_i = \cos\vartheta_i \hat{\bm{n}} + \sin \vartheta_i \bm{n}_\perp$. It is reasonable to assume that the pseudomonas bacteria mainly contributing to surface-mediated chemotaxis will have a pili distribution that is approximately symmetrical with respect to the body orientation (if this was not the case the bacterium would generate a torque in one preferred direction, rotate around itself and effectively stay in one point). Therefore, we can neglect terms in the sum over the pili that are odd in $\sin \vartheta_i$, like $\langle \sin \vartheta_i \rangle = \frac{1}{N} \sum_i \sin \vartheta_i$ or $\langle \cos \vartheta_i \sin \vartheta_i \rangle = \frac{1}{N} \sum_i \cos \vartheta_i \sin \vartheta_i$. The force can then be written as
\begin{equation}
\begin{split}
\bm{F} = &N \langle \cos \vartheta_i \rangle f(\psi) \bar{\Theta}(\psi) \hat{\bm{n}} + N \langle \sin^2 \vartheta_i \rangle \ell_p \partial_\psi \left[ f(\psi) \bar{\Theta}(\psi) \right] \bm{\nabla} \psi +  N \langle (1 - 2 \sin^2 \vartheta_i )  \rangle \ell_p \partial_\psi \left[ f(\psi) \bar{\Theta}(\psi) \right] \left(\bm{\nabla} \psi \cdot \hat{\bm{n}} \right) \hat{\bm{n}}  \\
&+ \hat{\bm{n}} \sum_i \cos \vartheta_i f(\psi_i) \delta \Theta_i + \hat{\bm{n}}_\perp \sum_i \sin \vartheta_i f(\psi_i) \delta \Theta_i.
\end{split}
\label{f-final}
\end{equation}
If the pili attachment events of two different pili are independent of each other we have $\langle \delta \Theta_i \delta \Theta_j \rangle = \sigma^2(\psi) \delta_{ij}$, where $\delta_{ij}$ is the Kronecker symbol. Moreover, we can calculate the auto-correlation of the attachment as $\langle \delta \Theta_i(t) \delta \Theta_i(t') \rangle = \sigma^2 \, e^{-(\mu+\lambda) |t-t'|} \,\delta_{ij}$. If we focus on time scales that are longer than the average attachment/detachment time, we can represent this auto-correlation as a delta function
\begin{equation}
\langle \delta \Theta_i(t) \delta \Theta_i(t') \rangle = \frac{\sigma^2(\psi)}{[\lambda(\psi) + \mu(\psi)]} \,\delta_{ij} \delta(t-t')=\frac{\lambda(\psi)\mu(\psi)}{[\lambda(\psi) + \mu(\psi)]^3}\,\delta_{ij} \delta(t-t').
\end{equation}
With this, we get for the mean square fluctuations parallel to $\hat{\bm{n}}$
\begin{equation}
%\begin{split}
\big\langle \big[ \hat{\bm{n}} \sum_i \cos \vartheta_i f(\psi_i) \delta \Theta_i \big]^2 \big\rangle  =  \sum_i \cos^2 \vartheta_i f^2(\psi_i) \sigma^2(\psi_i) =  N \langle \cos^2 \vartheta_i \rangle f^2(\psi) \sigma^2(\psi) + N \langle \cos^3 \vartheta_i \rangle \ell_p \partial_\psi \left[f^2(\psi) \sigma^2(\psi)\right] \left(\bm{\nabla} \psi \cdot \hat{\bm{n}}\right).
%\end{split}
\end{equation}
Analogously, for the mean square fluctuations perpendicular to $\hat{\bm{n}}_\perp$ we get
\begin{equation}
% \begin{split}
\big\langle \big[\hat{\bm{n}}_\perp \sum_i \sin \vartheta_i f(\psi_i) \delta \Theta_i \big]^2 \big\rangle   =  \sum_i \sin^2 \vartheta_i f^2(\psi_i) \sigma^2(\psi_i) =  N \langle \sin^2 \vartheta_i \rangle f^2(\psi) \sigma^2(\psi) + N \langle \sin^2 \vartheta_i \cos \vartheta_i \rangle \ell_p \partial_\psi \left[f^2(\psi) \sigma^2(\psi)\right] \left(\bm{\nabla} \psi \cdot \hat{\bm{n}}\right).
% \end{split}
\end{equation}
If $f(\psi)\sigma(\psi)$ only has a weak $\psi$-dependence or if the Psl concentration $\psi$ varies only little on the scale of a pilus length $\ell_p$, the gradient terms of order $| \bm{\nabla} \psi |$ in the variance are here negligible. But even if this is not the case, these terms will just slightly alter the fluctuation strengths in the parallel and perpendicular direction and are therefore not expected to introduce any new behavior. For these reasons we will neglect the gradient terms in the fluctuations but keep them in the deterministic part as they are responsible for new behavior.

The attachment/detachment of pili on the surface can be considered a set of $N$ random events and in case of $N \gg 1$ we can expect that the overall fluctuations in the force can be well approximated by a Gaussian with a variance of $N \langle \cos^2 \vartheta_i \rangle f^2(\psi) \sigma^2(\psi)$ in the direction $\hat{\bm{n}}$ and a variance of $N \langle \sin^2 \vartheta_i \rangle f^2(\psi) \sigma^2(\psi)$ in the direction $\hat{\bm{n}}_\perp$.

The pili-generated force will propel the bacterium and also generate a torque. In a viscous environment the center of mass velocity of a bacterium will be proportional to the force and therefore, we obtain the following translational equation of motion
\begin{equation}
%\begin{split}
\frac{d \bm{r}}{d t} = \frac{1}{\gamma_\parallel} \, \bm{F}_\parallel +  \frac{1}{\gamma_\perp} \, \bm{F}_\perp = v(\psi) \hat{\bm{n}} + A(\psi) (\bm{\nabla} \psi \cdot \hat{\bm{n}}_\perp) \hat{\bm{n}}_\perp + B(\psi) (\bm{\nabla} \psi \cdot \hat{\bm{n}}) \hat{\bm{n}} +  \sqrt{ 2 D_\parallel } \; \eta^\parallel \hat{\bm{n}} + \sqrt{2 D_\perp} \; \eta^{\perp} \hat{\bm{n}}_\perp,
\label{eq:position-nonoise}
%\end{split}
\end{equation}
where $\eta_\parallel$, $\eta_{\perp}$ are Gaussian noise components parallel and orthogonal to $\hat{\bm{n}}$ with $\langle \eta^{\parallel / \perp}\rangle = 0$, $\langle {\eta}^{\parallel}(t) {\eta}^{\parallel}(t')\rangle = \delta (t-t')$, $\langle {\eta}^{\perp}(t) {\eta}^{\perp}(t')\rangle =  \delta (t-t')$, and $\langle {\eta}^{\parallel}(t) {\eta}^{\perp}(t')\rangle = 0$.

This stochastic differential equation contains multiplicative noise and therefore needs an additional interpretation rule as discussed in \cite{van1981ito}: it could be interpreted in an It\^{o} or in a Stratonovich sense, or as a mix of the two. Phrasing the problem in It\^{o} notation \footnote{The It\^{o} notation implies evaluating all the terms at $\bm{r}$ but taking into account additional terms that enter if $\alpha \neq 0$ in $\bm{r} + \alpha d\bm{r}$.}, the confusion comes about if it is {\em a priori} unclear at which position $\bm{r} + \alpha d\bm{r}$ the noise should be evaluated ($\alpha = 0$ corresponds to the It\^{o} approach, whereas $\alpha = 1/2$ for Stratonovich). For Eq. \eqref{eq:position-nonoise} an interpretation with $\alpha \neq 0$ will in general modify the prefactors $A(\psi)$ and $B(\psi)$. Considering a general interpretation, in It\^{o} notation, the prefactors $v_0(\psi)$, $A(\psi)$, $B(\psi)$, $D_\parallel$ and $D_\perp$ are
\begin{eqnarray}
&&D_\parallel(\psi) = \frac{N}{2} \, \langle \cos^2 \vartheta_i \rangle \, \frac{f^2(\psi) }{\gamma_\parallel^2} \, \frac{\lambda(\psi)\mu(\psi) }{ [\lambda(\psi) + \mu(\psi)]^3 }, \\
&&D_\perp(\psi) = \frac{N}{2} \, \langle \sin^2 \vartheta_i \rangle \, \frac{f^2(\psi) }{\gamma_\perp^2} \, \frac{\lambda(\psi)\mu(\psi) }{ [\lambda(\psi) + \mu(\psi)]^3 }, \\
&&v(\psi) = N \, \langle \cos \vartheta_i \rangle \, \frac{f(\psi) }{\gamma_\parallel} \, \frac{\lambda(\psi)}{\lambda(\psi) + \mu(\psi)}, \\
&&A(\psi) = N \, \langle \sin^2 \vartheta_i \rangle \, \frac{\ell_p}{\gamma_\perp} \, \partial_\psi \left[ f(\psi) \, \frac{\lambda(\psi)}{\lambda(\psi) + \mu(\psi)} \right] + \alpha \partial_\psi D_\perp(\psi),\\
&&B(\psi) = N \, \langle \cos^2 \vartheta_i \rangle \, \frac{\ell_p}{\gamma_\parallel} \, \partial_\psi \left[ f(\psi) \, \frac{\lambda(\psi)}{\lambda(\psi) + \mu(\psi)}  \right] +  \alpha \partial_\psi D_\parallel(\psi).
\end{eqnarray}

The pulling force $\bm{F}$ in Eq. \eqref{f-final} will also generate a torque on the bacterial body. It is given by
\begin{equation}
 \bm{\tau} = \frac{\ell}{2} \, \hat{\bm{n}} \times \bm{F} =  \frac{\gamma \ell}{2} \, \hat{\bm{n}} \times \left[ A(\psi) \bm{\nabla} \psi +  \sqrt{2 D_\perp} \; \eta^{\perp} \hat{\bm{n}}_\perp \right].
\label{eq:torque-average}
\end{equation}
Note that the total resultant force is used as opposed to individual pili forces, because in our simplified model all the pili are attached to the same point.
On a surface in a viscous environment the rotational motion of the bacterium will be overdamped and therefore the angular velocity $\bm{\omega}$ will be linearly related to the torque.
\begin{equation}
\frac{d \hat{\bm{n}}}{d t} = - \hat{\bm{n}} \times \bm{\omega}  = - \frac{1}{\gamma_{\text{rot}}} \, \hat{\bm{n}} \times  \bm{\tau} =  -  \chi(\psi) \hat{\bm{n}} \times[\hat{\bm{n}} \times \bm{\nabla} \psi] +  \sqrt{2 D_r(\psi)} \; \eta^{\perp} \hat{\bm{n}}_\perp,
\label{n-final}
\end{equation}
where %$\bm{\eta}_r = \eta_\perp \hat{\bm{e}}_z$,
\begin{equation}
\chi(\psi) = \frac{\ell \ell_p}{2 \gamma_{\text{rot}}} N \, \langle \sin^2 \vartheta_i \rangle \partial_\psi \left[f(\psi) \, \frac{\lambda(\psi)}{\lambda(\psi) + \mu(\psi)} \right] +  \alpha   \partial_\psi D_r,
\end{equation}
and
\begin{equation}
D_r =  \left(\frac{\gamma_\perp \ell}{2 \gamma_{\text{rot}}} \right)^2 D_\perp,
\end{equation}
taking into account a general interpretation rule $\alpha \neq 0$ for the multiplicative noise.

We can now rewrite the above expressions in terms of $v(\psi)$ and $D_r(\psi)$ in the following way
\begin{eqnarray}
&&D_\parallel(\psi) = \left[\frac{4 \gamma_{\text{rot}}^2  c_2}{  \ell^2 \gamma_\parallel^2 (1-c_2)}\right] D_r(\psi), \\
&&D_\perp(\psi) = \left( \frac{2 \gamma_{\text{rot}} }{ \gamma_\perp \ell } \right)^2 D_r(\psi),\\
&&\chi(\psi) = \left[\frac{\gamma_{\parallel} \ell \ell_p (1-c_2)}{2 \gamma_{\text{rot}} c_1} \right] \partial_\psi v(\psi)  +  \alpha  \partial_\psi D_r(\psi),\\
&&A(\psi) =  \left[\frac{\ell_p (1-c_2) \gamma_\parallel}{c_1 \gamma_\perp} \right] \partial_\psi v(\psi)  + \alpha \left( \frac{2 \gamma_{\text{rot}} }{ \gamma_\perp \ell } \right)^2 \, \partial_\psi D_r(\psi),\\
&&B(\psi) =   \left[\frac{\ell_p c_2}{c_1}\right] \partial_\psi v(\psi) + \alpha \left[\frac{ 4 \gamma_{\text{rot}}^2 c_2}{\ell^2 \gamma_\parallel^2 (1-c_2)  }\right] \partial_\psi D_r(\psi),
\end{eqnarray}
where $c_1 = \langle \cos \vartheta_i \rangle$ and $c_2=\langle \cos^2 \vartheta_i \rangle$ are assumed to be Psl independent.

\begin{figure}[b]
%\centering
\includegraphics[width = 0.99\columnwidth]{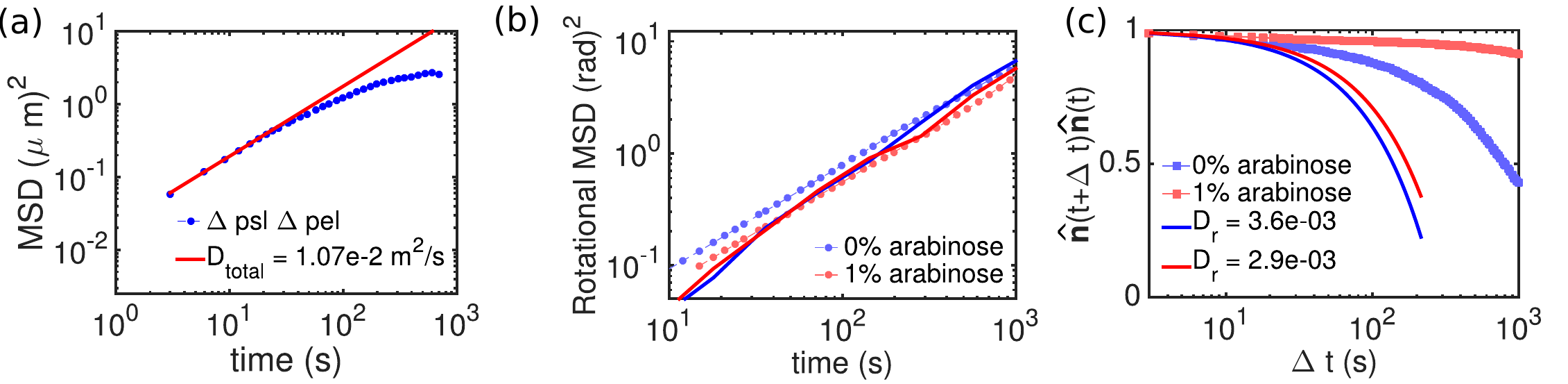}
\caption{(a) Effective translational MSD for the $\Delta psl \Delta pel$ mutant which does not secrete EPS. For short times we see purely diffusive behaviour with a diffusion constant $D_{\text{total} 0} =   D_{\parallel 0} + D_{\perp 0} \approx 1.07\times10^{-2} \mu \text{m}^2\,\text{s}^{-1}$. For longer times moving bacteria will detach from the surface, such that only non-motile bacteria will contribute to the MSD. (b) Rotational MSD for the $\Delta$P$_{psl}$/P$_{BAD}$-{\it psl} mutant under $0\%$ arabinose (blue dots) and  $1\%$ arabinose (red dots), as well as a comparison with corresponding simulations ($k = 0.073\,\text{s}^{-1}$ for $0\%$ arabinose and $k = 0.18\,\text{s}^{-1}$ for $1\%$ arabinose). From this we determined the rotational diffusion constant $D_r \approx 3.6 \times 10^{-3}\,\text{s}^{-1}$, noting that $D_r$ is only weakly dependent on $\psi$. (c) Extracting $D_r$ from the orientational correlation $\langle \hat{\bm{n}}(t + \Delta t) \hat{\bm{n}}(t) \rangle$ (thick points), and the short time approximation $1 -\langle  \left[\varphi(t + \Delta t) - \varphi(t) \right]^2 \rangle /2$ (thin lines). They yield similar values for $D_r$. \label{fig:alphamsds}}
\end{figure}

\section{\fontsize{12}{15} \sc \selectfont III. Extracting the Parameters from the Experiments}

Here we provide some of the details involved in extracting single-bacteria parameters from the experimental observations. To extract the translational diffusion coefficient, we track a mutant which does not secrete EPS, $\Delta psl \Delta pel$, and find a diffusion constant $D_{\text{total} 0} = 1.1 \times10^{-2} \mu \text{m}^2\,\text{s}^{-1}$ [see Fig. S2(a)]. This value gives the sum of parallel and perpendicular diffusion coefficients $D_{\text{total}0} = D_{\parallel 0} + D_{\perp 0} = D_{\perp 0} [1 + c_2 \gamma_\perp^2 / (1-c_2) \gamma_\parallel^2]$. We note that the experimental data for the translational MSD shown in Fig. S2(a) shows no propulsive component of the motion at the shortest timescale that we have resolved. We thus conclude that in the absence of Psl the velocity component $v_0 \approx 0$, and we can model the velocity as $v = v_1 \psi$.

The rotational MSD curves are shown in Fig. S2(b) for $0 \%$ and $1 \%$ arabinose, and comparison with our simulations. The rotational diffusion constant appears to depend on $\psi$ relatively weakly, and thus we use an approximation in which we assign the value of $D_r \approx 3.6 \times 10^{-3}\,\text{s}^{-1}$ in the entire range of arabinose concentration. An alternative way to determine $D_r$ from the orientational correlation $\langle \hat{\bm{n}}(t + \Delta t) \hat{\bm{n}}(t) \rangle = \langle \cos(\varphi(t + \Delta t) - \varphi(t)) \rangle$, which is approximately $1 -\langle  \left[\varphi(t + \Delta t) - \varphi(t) \right]^2 \rangle /2$, yields similar values for $D_r$; see Fig. S2(c). Note that we need to focus only on the first few time points to extract the intrinsic value of $D_r$. At longer times, the director auto-correlation function shows a crossover to a regime where the orientation is randomized less strongly overtime, which is a signature of  the stabilization of the orientation. 

Noting that the rotational diffusion $D_r$ varies only weakly between $0 \%$ and $1 \%$ arabinose, we set $D_{r 1} \approx 0$, which also introduces non-zero translational diffusion constants $D_\parallel$ and $D_\perp$. With $D_{r} \approx 3.6 \times10^{-3} \,\text{s}^{-1}$, and our measurement for the overall translational diffusion coefficient, we get $D_{\perp} = 2.3\times10^{-3} \mu \text{m}^2\,\text{s}^{-1}$ and $c_2 = \langle \cos^2 \vartheta \rangle \approx 0.7$. As an approximation, in our simulations we will set $c_1 = \langle \cos \vartheta \rangle \approx \sqrt{c_2} = 0.8$.

%\anote{From experiments we found that $D_r(\psi)$ varies only little with EPS secretion $k$ compared to other parameters [see Fig. \ref{fig:alphamsds}(b)--\ref{fig:alphamsds}(c)] and hence treated $D_r$ as a constant in our simulations.}
%We found that the correction due to the ambiguity in the multiplicative noise does not significantly influence our results, and hence chose $\alpha = 0$ for the results that we have presented in the main paper, for simplicity.

%\newpage

\section{\fontsize{12}{15} \sc \selectfont IV. Notes on Trail Alignment}

%While \emph{P. aeruginosa} do not necessarily reorient themselves tangentially to an existing track \cite{Zhao:2013}, alignment and trail-following are an important manifestation of the preferential association with Psl-rich regions on a surface.

\begin{figure}[b]
%\centering
\includegraphics[width =\textwidth]{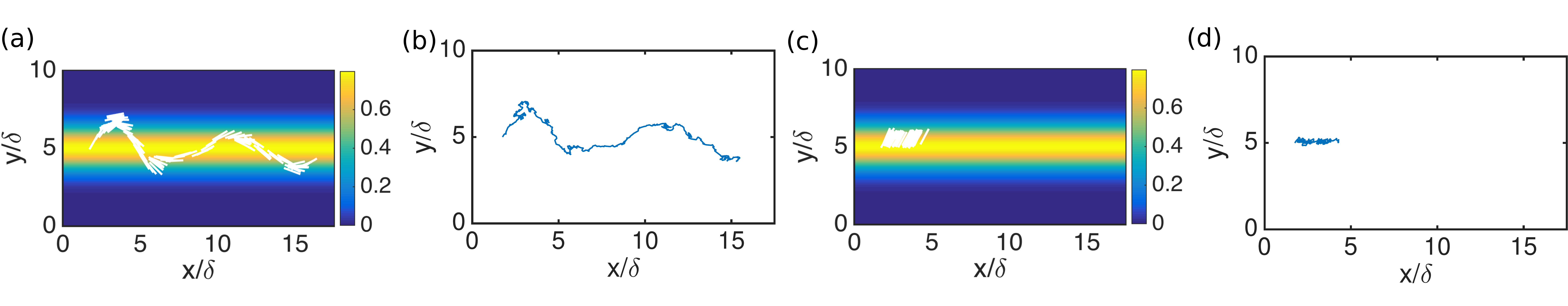}
\caption{Typical behavior of a bacterium around an existing Gaussian Psl trail (color-coded), located at $y=5 \delta$, after $t = 84$ s. All parameters are the same as in the main text, with $k = 0.13 \,\text{s}^{-1}$. Snapshots of bacterial positions and orientations (white) were taken every $0.15$ s. For better presentation the size of the bacteria is not to scale, the thickness of a real bacterium is of the order of the trail width $\delta$.In (a) and (b) we set parallel and perpendicular displacement terms proportional to $A(\psi)$ and $B(\psi)$ to zero. In (c) and (d) we set the $\chi(\psi)$ to zero and keep the anisotropic terms $A(\psi)$ and $B(\psi)$. (a) Starting from an orientation of $30^{\circ}$ away from the trail (left), the bacterium follows the trail in an oscillatory manner (b) shows the bacterial trajectory and oscillations along the trail. (c) and (d) shows a typical trajectory without the orientational alignment term but with the translational and rotational displacement terms proportional to $A(\psi)$ and $B(\psi)$. The bacterium can move to the center of the trail but trail following is decreased because the bacterium stays at a $30^{\circ}$ angle relative to the trail and cannot reorient.  \label{existingTrailAlignment}}
\end{figure}

In this section we will briefly discuss the mechanisms how the model derived above can lead to trail following. We consider the model given by equations \eqref{eq:position-nonoise} and \eqref{n-final} in a geometry where a bacterium interacts with an existing trail. This trail-following behavior is due to the displacement terms $A(\psi) (\bm{\nabla} \psi \cdot \hat{\bm{n}}_\perp) \hat{\bm{n}}_\perp$ and $B(\psi) (\bm{\nabla} \psi \cdot \hat{\bm{n}}) \hat{\bm{n}}$ in Eq. \ref{eq:position-nonoise}, which try to maximize the local Psl concentration by moving up the Psl gradient. However, more counter-intuitively, trail-following is also made possible by the orientational term $\chi(\psi) \hat{\bm{n}} \times[\bm{n} \times \bm{\nabla} \psi] $ in Eq. \eqref{n-final}. The reason for this is that the term reorients the bacterium in the direction of the Psl gradient, i.e towards the inside of a trail. Once a bacterium passes the center of the existing trail, the gradient switches signs and the alignment term will reorient the bacterium towards the inside of the trail again. The result is oscillatory trail-following.

To illustrate the influence of the orientational alignment term we have set the gradient displacement terms proportional to $A(\psi)$ and $B(\psi)$ to zero and kept all the other parameters as in the main paper. A typical result can be seen in Figs. S3(a) and S3(b), which shows oscillations and trail following. On the other hand, in the presence of gradient displacement terms proportional to $A(\psi)$ and $B(\psi)$ and absence of the gradient alignment term, we observe a different behavior shown in Figs. S3(c)--S3(d). While the gradient terms will move the bacterium to the center of the trail, the lack of reorientation does not allow efficient trail-following.

The combination of gradient-dependent displacement and alignment with the gradient constitutes a robust mechanism for  the attraction of bacteria to existing Psl trails and for the movement along Psl-rich regions.

\twocolumngrid

\end{document}